\documentclass[a4paper,showpacs,amsmath,twocolumn,amssymb,floatfix,prl,preprintnumbers,superscriptaddress,footinbib]{revtex4}
\usepackage[dvips]{graphicx}
\usepackage{dcolumn}
\usepackage{bm}
 \usepackage{latexsym}
 \usepackage{amsmath}
 \usepackage{amsthm}
 \usepackage{amsfonts}
 \usepackage{epsfig}
 \usepackage{graphics}

\begin{document}


\title{All-electrical detection of the relative strength of Rashba and Dresselhaus \\
     spin-orbit interaction in quantum wires}

\author{Matthias Scheid}
\email{Matthias.Scheid@physik.uni-r.de}
\affiliation{Graduate School of Engineering, Tohoku University, 6-6-02 Aramaki-Aza Aoba, Aoba-ku, Sendai 980-8579, Japan}
\affiliation{Institut f\"ur Theoretische Physik, Universit\"at Regensburg, 93040 Regensburg, Germany}
\author{Makoto Kohda}
\affiliation{Graduate School of Engineering, Tohoku University, 6-6-02 Aramaki-Aza Aoba, Aoba-ku, Sendai 980-8579, Japan}
\author{Yoji Kunihashi}
\affiliation{Graduate School of Engineering, Tohoku University, 6-6-02 Aramaki-Aza Aoba, Aoba-ku, Sendai 980-8579, Japan}
\author{Klaus Richter}
\affiliation{Institut f\"ur Theoretische Physik, Universit\"at Regensburg, 93040 Regensburg, Germany}
\author{Junsaku Nitta}
\affiliation{Graduate School of Engineering, Tohoku University, 6-6-02 Aramaki-Aza Aoba, Aoba-ku, Sendai 980-8579, Japan}

\begin{abstract}
We propose a method to determine the relative strength of Rashba and Dresselhaus spin-orbit
interaction from transport measurements without the need of fitting parameters. To this end, we make use of
the conductance anisotropy in narrow quantum wires with respect to the directions of an in-plane
magnetic field, the quantum wire and the crystal orientation. We support our proposal by numerical
calculations of the conductance of quantum wires based on the Landauer formalism which show the applicability of the method to a wide range of parameters.
\end{abstract}

\maketitle

With conventional electronics expected to reach critical boundaries for its performance soon, 
a new field of research utilizing the spin of the electron has evolved in recent years. Within this field
called spintronics much attention has been focussed on spin-orbit interaction (SOI) because it
provides a way of controlling the spin degree of freedom electrically in (non-magnetic) semiconductor-based 
systems without the need of external magnetic fields. However, SOI in two-dimensional electron gases
(2DEG) is a double-edged sword, since spin relaxation in disordered 2DEGs, which is typically
dominated by the D'yakonov-Perel' (DP) mechanism~\cite{Dyakonov1971}, is enhanced for strong SOI.
Since many promising semiconductor spintronics device proposals, e.g. the Datta-Das spin field effect
transistor (SFET)~\cite{Datta1990}, rely on coherent spin transport, it is desirable to efficiently
suppress the spin relaxation. In 2DEGs formed in III-V semiconductor heterostructures, there are
typically two main SOI contributions, namely, Rashba SOI due to structural inversion
asymmetry~\cite{Rashba1960} and Dresselhaus SOI due to bulk inversion asymmetry of the semiconductor
crystal~\cite{Dresselhaus1955}. An interesting situation occurs when the $k$-linear Rashba and
Dresselhaus terms are of equal strength, {\em i.e.}~$\alpha\!=\!\beta $. 
Then, spin is a good quantum number and DP spin
relaxation is absent~\cite{Schliemann2003}. Lately there has been much effort into this direction both theoretically with
new device proposals~\cite{Schliemann2003,Cartoixa2003}, and experimentally with the aim to achieve $\alpha =\beta $~\cite{Giglberger2007}.
Naturally, a precise control of the ratio $\alpha /\beta$ is essential for spin manipulation and the
operability of many spintronics devices. Since the strength $\beta$ of the Dresselhaus SOI is fixed in
a given quantum well the most promising tool to modify $\alpha /\beta$ is the control of the Rashba
SOI strength $\alpha$ via gate voltages~\cite{Nitta1997}.

To operate spintronics setups relying on the value of $\alpha /\beta$ requires 
the ability to measure this ratio with high accuracy. Although it is possible to determine $\alpha /\beta$ 
by using optical techniques~\cite{Giglberger2007,Ganichev2004,Meier2007}, this is not always an option. 
If, e.g., the semiconductor heterostructure is covered by a top gate used to tune the Rashba SOI
strength, it is very difficult to carry out optical measurements; therefore methods are highly desirable that
allow one to determine the ratio  $\alpha /\beta$ from transport measurements.
In principle, this can be achieved by fitting weak antilocalization (WAL) data from magneto-conductance (MC)
measurements to analytical predictions~\cite{Knap1996,Miller2003}. However, the results usually bear a 
certain ambiguity, since one has to fit the data with several parameters and the possible error 
margins are thus quite large.

Hence, in this Letter we propose an alternative, all-electrical method to determine 
the relative strength, $\alpha /\beta$, of Rashba and Dresselhaus SOI from measuring the conductance 
of narrow quantum wires defined in a 2DEG subject to an in-plane magnetic field. The method is based 
on the fact, that only for a field 
parallel to the effective magnetic field due to SOI the weak localization (WL) correction to the
conductance survives, while it is suppressed for all other directions. No fit parameters 
are required, and $\alpha /\beta$ is straightforwardly related to this specific field direction,
where the conductance is minimal.

We numerically calculate the conductance $G$ of a disordered quantum wire realized in a 2DEG with SOI
linear in momentum. The single-particle Hamiltonian of the quantum wire in $x$-direction 
reads~\cite{Lusakowski2003}
\begin{equation}\label{Ham}
\mathcal{H}=\frac{\pi_x^2+\pi_y^2}{2m^*} + U(x,y) + \frac{\mu _\text{B}g^*}{2}\big(\vec{B}_{||}+\vec{B}_\text{so}(\vec{\pi})\big)\cdot\vec{\sigma},
\end{equation}
with the effective spin-orbit field
\begin{eqnarray}\label{Bso}
\vec{B}_\text{so}(\vec{\pi})&=\frac{2}{\mu _\text{B}g^*\hbar}&\Big[\hat e _x \big( \alpha\pi_y+\beta(\pi_x\cos{2\phi}-\pi_y\sin{2\phi})\big)\\
&\qquad\quad+&\hat e _y \big( -\alpha\pi_x-\beta(\pi_x\sin{2\phi}+\pi_y\cos{2\phi})\big)\Big] \nonumber
\end{eqnarray}
and the external in-plane magnetic field 
\begin{equation}\label{Bin}
\vec{B}_{||}=B_{||} (\cos (\theta -\phi ) \hat e_x + \sin (\theta -\phi ) \hat e_y ).
\end{equation}
The vector potential components $A_{i}$ in $\pi_{i}=(p_{i}+eA_{i})$ arise due to the perpendicular magnetic field $B_z$ whose contribution to the Zeeman effect we neglect. In Eq.~(\ref{Bso}) $\alpha$ and $\beta$ is the Rashba and Dresselhaus SOI strength respectively and $\phi$/$\theta$ is the angle between the quantum wire/in-plane magnetic field and the $[100]$ direction of the crystal for a zinc-blende heterostructure grown in the $[001]$ direction. The electrostatic potential $U(x,y)$ includes the confining potential for the quantum wire and the disorder potential from static non-magnetic impurities in a region of length $L$.
For the calculations we use a discretized version of the Hamiltonian~(\ref{Ham}) that allows us to evaluate the transport properties of the wire by computing lattice Green functions. For details see, e.g., Ref.~\cite{Wimmer2008}.\\
The dimensionless numerical parameters used in this letter (denoted by a bar) are related to real
physical quantities as follows (for square lattice spacing $a$):  Energy $\bar{E} =(2m^*a^2/\hbar ^2)E$, SOI strengths $\bar{\alpha}=(m^*a/\hbar ^2)\alpha$ and $\bar{\beta}=(m^*a/\hbar ^2)\beta$. As a typical lengthscale for the simulations we introduce $W_0=20a$. In the calculations, the disorder potential is modelled by Anderson disorder with strength $\bar{U}_0$. The mean free path is given by
$l=2.4W_0\sqrt{\bar{E_\text{F}}}/\bar{U}_0^2$, where $\bar{E_\text{F}}$ is the scaled Fermi energy. The conductance 
of the wire is obtained by averaging over $N_\text{d}$ disorder configurations and unless stated otherwise 
the following parameters are fixed : $\bar{E}_\text{F}=0.5$ (corresponding to 4 propagating modes for a wire of width $W_0$), $L=7.5W_0$,
$\bar{U}_0=1.4$ (i.e. $l\approx 0.87W_0$) and $N_\text{d}=10000$. 

\begin{figure}[tb!]
	\centering 
	\includegraphics[width=\linewidth]{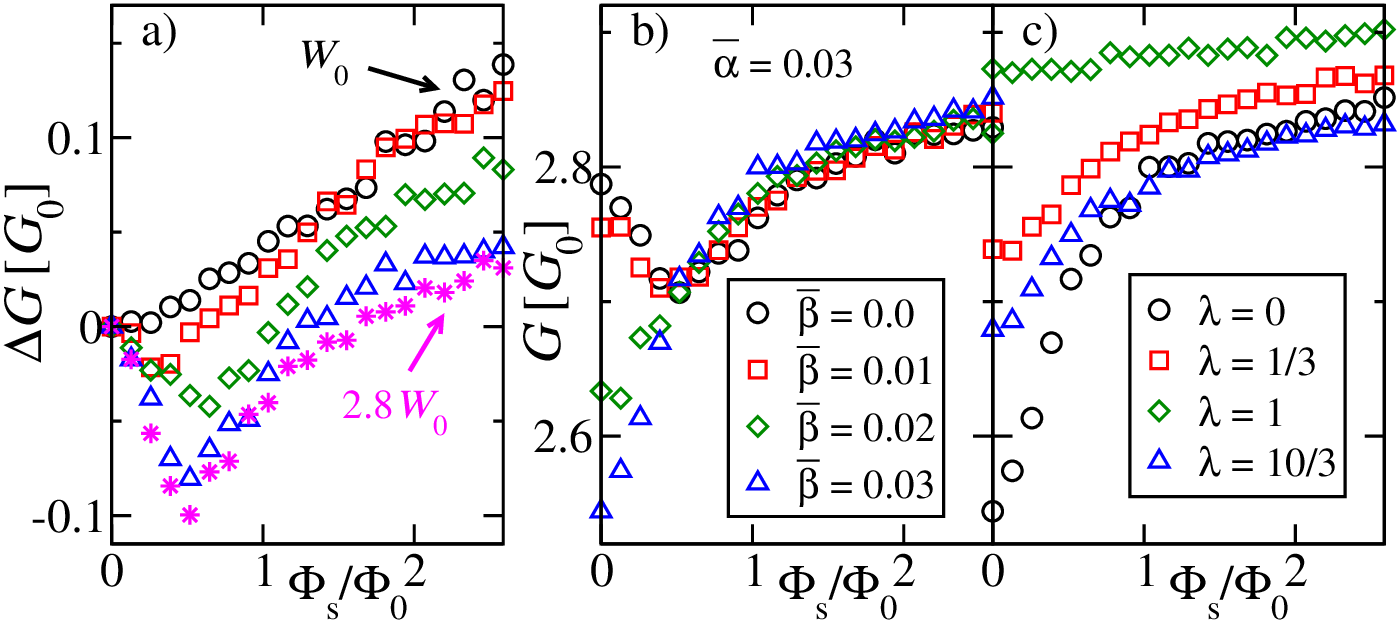}
	\caption{(\emph{Color online}) Magnetoconductance of a quantum wire plotted against the magnetic flux $\Phi _s=W_0^2B_z$ in units of $\Phi _0=h/e$. a) $\Delta G= G(\Phi _s)-G(0)$ for $\bar{\alpha}=0.03$ (i.e. $L^{\alpha}_\text{SO}\approx 5.2W_0$), $\bar{\beta}=0.0$ and widths $W=W_0$, $1.3W_0$, $1.8W_0$, $2.3W_0$, $2.8W_0$ from top to bottom. b) $\bar{\alpha}=0.03$, $W=2.3W_0$, $\phi=\pi /2$ for several values of $\bar{\beta}$.  c) $\bar{\alpha}=\bar{\beta}=0.03$, $W=2.3W_0$, $\phi=\pi /2$ and $\theta =\pi$ for several values of $\lambda$.}
	\label{Fig1}
\end{figure}

\begin{figure}[tb!]
	\centering 
	\includegraphics[width=\linewidth]{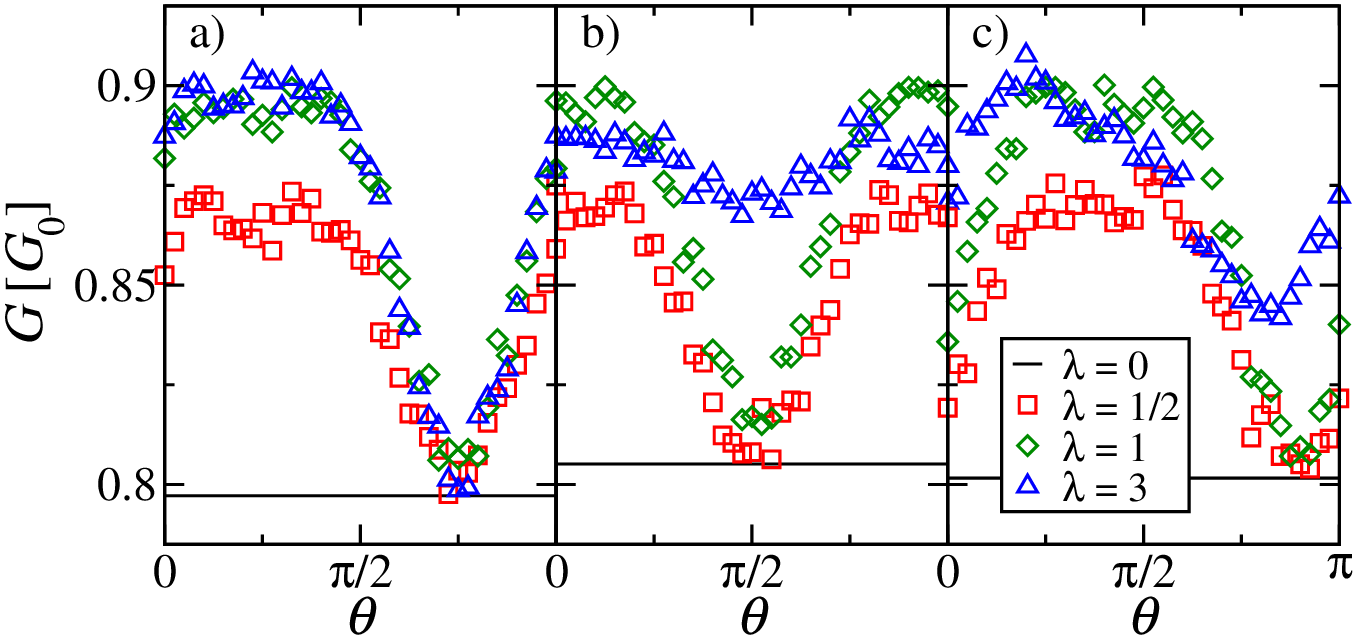}
	\caption{(\emph{Color online}) Conductance of a wire of width $W_0$ at $\Phi _s=0$, $\phi=\pi /2$ and fixed $(\mu _\text{B}g^*m^*a^2/\hbar ^2)|\vec{B}_\text{so}(k_x)|= \sqrt{2(\bar{\alpha} ^2 +\bar{\beta} ^2)}=0.02$ with respect to $\theta$ for different values of $\lambda$. a) $\bar{\alpha}=\bar{\beta}$ b) $\bar{\alpha} =0$ c) $\bar{\alpha}=3\bar{\beta}$.}
	\label{Fig3}
\end{figure}

To understand the mechanism for the detection of $\alpha /\beta$, which requires finite $\vec{B}_{||}$, we first study the conductance of quantum wires at $B_{||}=0$. Specifically we present the MC for two cases, where WAL is suppressed: (a) Rashba and Dresselhaus spin precession lengths larger than the width of the wire $W$, i.e., $L^{\alpha}_\text{SO}= (\pi\hbar ^2/m^*\alpha ) \gg W$, $L^{\beta}_\text{SO}= (\pi\hbar ^2/m^*\beta ) \gg W$ and (b) $\alpha =\beta$.\\
In Fig.~\ref{Fig1}a, we plot $G(\Phi _s)-G(0)$ for wires with fixed $\alpha\neq 0, \beta =0$ and different widths $W$, showing that for smaller $W$ WAL is suppressed, which is in line with earlier experimental results~\cite{Schapers2006} and confirms  analytical~\cite{Kettemann2007} and numerical treatments~\cite{Schapers2006}. Since spin relaxation is essential for WAL, the mechanism for the suppression of WAL can be attributed to an enhancement of the spin-scattering length in narrow wires~\cite{Kiselev2000,Holleitner2006}, and more generally, in confined geometries~\cite{Aleiner2001,Zaitsev2005}.\\ 
In the case (b), $\alpha =\beta$, $\vec{B}_\text{so}$ points uniformly into the [$\bar{1}10 $]-direction for all $k$-vectors and a so-called persistent spin helix forms~\cite{Bernevig2006}. There the spin state of an electron is determined only by its initial and final position independent of the exact path in-between. Therefore, charge carriers do not acquire an additional phase due to SOI upon return to their initial positions, resulting in constructive interference of the wavefunctions connected by time reversal, hence WL~\cite{Pikus1995}. This behavior is shown for fixed $W$ and $\alpha$ but variable $\beta$ in Fig.~\ref{Fig1}b where we observe that WAL is suppressed for $\alpha =\beta$.\\
In both cases shown in Figs.~\ref{Fig1}a,b the absence of WAL is caused by the suppression of spin relaxation with the spin relaxation length exceeding the length of the wire $L$, where $L$ in the numerical simulation takes the role of the phase coherence length in the experiment.
\\
We now investigate the influence of an additional in-plane magnetic field on the conductance of a quantum wire
where WAL is suppressed. For convenience, we introduce the ratio $\lambda =B_{||}/|\vec{B}_\text{so}(k_x)|$ which is the relative strength of the in-plane magnetic field and the effective magnetic field due to SOI for a $k$-vector along the quantum wire, see Eqs.~(\ref{Bso}),(\ref{Bin}). In Fig.~\ref{Fig1}c we show the MC for the case $\alpha =\beta$ for several values of $\lambda$: The conductance at $\Phi _s =0$ is enhanced by a finite $B_{||}$. The form of the MC curves in Fig.~\ref{Fig1}c can be understood from the expression for the WL/WAL conductance correction from diagrammatic perturbation theory~\cite{Hikami1980}. It is of the form $\Delta G\propto (C_{00}-\sum^{1}_{m=-1} C_{1m})$, where the first (singlet) term $C_{00}$ contributes positively to the conductance and is responsible for the typical WAL peak in systems with SOI. It is unaffected by DP spin
relaxation but suppressed by an in-plane magnetic field~\cite{Malshukov1997}. 
The second (triplet) term gives a negative conductance
contribution and is suppressed for short spin relaxation times~\cite{Hikami1980}. For the parameters used
in Fig.~\ref{Fig1}c, $C_{00}$ is suppressed for $\lambda\ge 0.15$, thus in the respective curves shown in Fig.~\ref{Fig1}c only the triplet term is present in $\Delta G$ resulting in positive MC $(\partial G/\partial \Phi _s)>0$. While for $\lambda =0$ we observe WL due to $\alpha =\beta$, increasing $\lambda$ gives rise to a transition to $\partial G/\partial \Phi _s \approx 0$ at $\lambda \approx 1$ and back to WL for $\lambda \gg 1$. This can be understood by the change of the spin relaxation in the system: For finite $\vec{B}_{||}$ in a direction different from $[ \bar{1}10 ]$ ($\theta=3\pi /4$), the resulting magnetic field $\vec{B}_\text{tot}(\vec{\pi})=\vec{B}_{||}+\vec{B}_\text{so}(\vec{\pi})$ will not be uniformly in the $[ \bar{1}10 ]$ direction anymore, but cause spin relaxation, which is strongest for comparable strengths of $\vec{B}_{||}$ and $\vec{B}_\text{so}$ and yields a reduction of the triplet term (green diamonds in Fig.~\ref{Fig1}c). 
For in-plane magnetic fields which distinctly exceed the effective magnetic field ($\lambda\gg 1$), on the other hand, WL is restored to some degree (blue triangles in Fig.~\ref{Fig1}c) since the resulting $\vec{B}_\text{tot}(\vec{\pi})$ is strongly aligned in the direction of $\vec{B}_{||}$ and spin relaxation is reduced again. The enhancement of $G(\Phi _s=0)$ in an in-plane magnetic field is anisotropic with respect to the direction of $\vec{B}_{||}$. For $\theta =(3/4)\pi$, spin remains a good quantum number due to $\vec{B}_{||}\parallel\vec{B}_\text{so}$. Thus DP spin relaxation is absent, resulting in WL. This behavior can be observed in Fig.~\ref{Fig3}a, where $G(\theta)$ at $\Phi _s =0$ is shown for a slightly different geometry. 
Contrary to the case considered here, in systems showing WAL for $B_{||}=0$, the transition from WAL to WL is observed with increasing $B_{||}$~\cite{Meijer2004,Meijer2005} due to the reduction of the singlet term  caused by $\vec{B}_{||}$.\\
We now investigate the conductance subject to an in-plane magnetic field in quantum wires where WAL is suppressed due to a much smaller width with respect to the spin precession lengths. 
In Fig.~\ref{Fig3} we plot the dependence of the conductance on the angle $\theta$ for three different ratios $\alpha /\beta$. In order to understand the increase of $G$ at $\lambda>0$ for all but one angle $\theta$, we consider the case of a strictly one-dimensional quantum wire (1DQW) with SOI. We follow this approach, since for the system investigated in Fig.~\ref{Fig3} the width of the wire is much smaller than the phase coherence length, a situation where it is sufficient to take into account only the transversal zero-mode for the calculation of the quantum correction to the conductance~\cite{Kettemann2007}. A disordered 1DQW exhibits WL even if SOI of the Rashba and/or Dresselhaus type is present, since the spin is a conserved quantity in this limit. The effective magnetic field experienced by the
electrons is exactly opposite for electrons travelling in $+\hat x$ or $-\hat x$-direction, and thus
no additional phase in the wavefunction is acquired by electrons returning to their original
position. However, a finite in-plane magnetic field can suppress the WL and induce an increase in the conductance. If $\vec{B}_{||}\nparallel \vec{B}_\text{so}(k_x)$, the direction of the total magnetic field, $\vec{B}_{||}+\vec{B}_\text{so}(k_x)$, is different for electrons travelling in $+\hat x$ or $-\hat x$-direction, resulting in spin relaxation and an increase of $G$ (reduction of WL). 
A minimum in $G(\theta )$ exists for $\vec{B}_{||}\parallel \vec{B}_\text{so}(k_x)$, where no DP spin relaxation takes place since spin is still a good quantum number. 
In Fig.~\ref{Fig3} we observe that the minimum of $G$ appears at the angle which corresponds to the respective effective magnetic field direction for a $k$-vector along the wire direction.

In view of the results of Fig.~\ref{Fig3}, we conjecture that also for a quasi-one-dimensional quantum wire with $W\ll L^{\alpha /\beta}_\text{SO}$ the angle at which the minimum in the conductance appears is given by the direction of the effective magnetic field $\vec{B}_\text{so}(k_x)$ for a $k$-vector along the wire direction $\hat{x}$:
\begin{equation}\label{fullthetamin}
\theta _\text{min} =\arctan\left(-\frac{\alpha\cos\phi+\beta\sin\phi}{\beta\cos\phi+\alpha\sin\phi}\right).\\
\end{equation}
\begin{figure}[tb!]
	\centering 
	\includegraphics[width=0.8\linewidth]{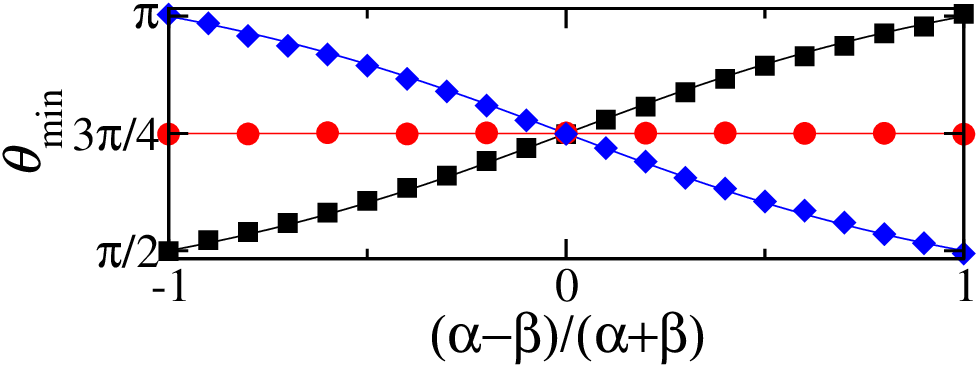}
	\caption{(\emph{Color online}) $\theta_\text{min}$ determined numerically for a system with $W_0$, $\bar{\alpha} +\bar{\beta} =0.04$, $\bar{U}_0=1.2$, $(\mu _\text{B}g^*m^*a^2/\hbar ^2)B_{||}=0.01$ and $N_\text{d}=20000$. Black squares: $\phi =\pi /2$; red circles: $\phi = \pi /4$; blue diamonds: $\phi = 0$. The solid lines represent Eq.~(\ref{fullthetamin}) for the respective angles $\phi$.} 
	\label{Fig5}
\end{figure}
In Fig.~\ref{Fig5}, we plot Eq.~(\ref{fullthetamin}) for three different wire orientations $\phi$ (solid lines), whose validity is nicely confirmed by extracting $\theta _\text{min}$ from the numerical $G(\theta )$ dependence (such as Fig.~\ref{Fig3}) for different ratios of $\alpha /\beta$ (symbols) with fixed $\alpha +\beta$. In order to use this feature for the determination of the ratio $\alpha /\beta$ we suggest to measure $G(\theta )$ for quantum wires oriented either along the [100] or the [010]-direction. Then the angle of the minimum conductance directly provides the unambiguous value for the relative strength and signs of $\alpha$ and $\beta$. Choosing, e.g. $\phi =\pi /2$ this ratio is given by $\alpha /\beta =-\cot (\theta _\text{min})$, which is representative for the whole sample, since the influence of the lateral confinement on the strength of the SOI is negligible~\cite{Guzenko2006}. Considering quantum wires realized in an InAlAs/InGaAs heterostructure (typical values $m^*=0.05m_0$, $g^*=3$) and fixing the width $W_0=350$nm, we see that the parameters used in Fig.~\ref{Fig5} ($l\approx 412$nm, $B_{||}\approx 0.17$T and $\alpha +\beta \approx 3.5\cdot 10^{-12}e$Vm) are well in reach of present day experiments~\cite{Meijer2004,Bergsten2006}.\\
\begin{figure}[tb!]
	\centering 
	\includegraphics[width=0.8\linewidth]{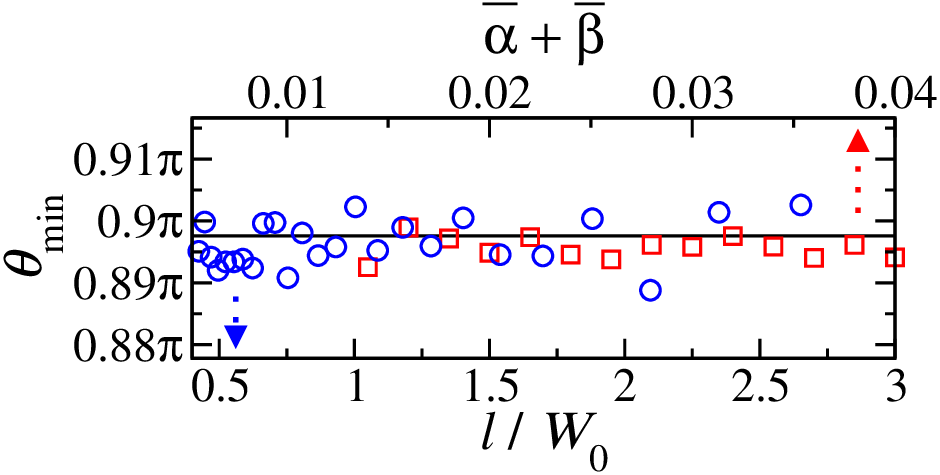}
	\caption{(\emph{Color online}) Numerically determined $\theta_\text{min}$ for $W_0$, $\phi =\pi /2$, $(\mu _\text{B}g^*m^*a^2/\hbar ^2)B_{||}=0.01$, $N_\text{d}=8000$ and $\alpha /\beta =3$. Either the mean free path $l$ for fixed $\bar{\alpha} +\bar{\beta}=0.02$ (blue circles) or $\bar{\alpha} +\bar{\beta}$ for fixed $l\approx 0.87W_0$ (red squares) was varied. The black line shows the expected value of $\theta_\text{min}$ from Eq.~(\ref{fullthetamin}).} 
	\label{Fig6}
\end{figure}
We have neglected effects due to the cubic Dresselhaus SOI term, which becomes increasingly important for 
wide quantum wells. In general, it induces additional randomization of the spin state, which for the
case of a very strong cubic Dresselhaus contribution can result in the absence of the suppression of
WAL~\cite{Pikus1995}. Nevertheless, since cubic Dresselhaus coupling is smallest for $k$-vectors along [100] or
[010] directions, we have neglected it for the determination of $\alpha /\beta$, since in our proposal the quantum wire is
assumed to be oriented in one of those directions. However, in contrast to a 1DQW, it might have an effect on $\theta_\text{min}$, if it is comparable in strength to the linear term.

In order to assess possible limitations of this method, we performed calculations varying several parameters, while keeping the ratio $\alpha /\beta =3$ constant. In Fig.~\ref{Fig6}, we show that Eq.~(\ref{fullthetamin}), $\theta_\text{min}=\arctan(-1/3)\approx 0.9\pi$,  is fulfilled for a wide range of both SOI strengths (squares) and mean free paths (circles). Further numerical calculations, upon increasing the number of transverse orbital modes in the wire up to 13, showed that Eq.~(\ref{fullthetamin}) still holds true (not presented here).\\
In conclusion we have shown, that Eq.~(\ref{fullthetamin}), derived for a 1DQW, provides a 
valuable tool to determine the ratio $\alpha /\beta$ also for a quantum wire with several transversal 
modes, only requiring $W\ll L^{\alpha /\beta}_\text{SO}$, i.e. a suppression of WAL due to the 
confinement~\cite{Schapers2006}. For increasing width, $G(\theta )$ evolves into a behavior 
typical of a 2DEG~\cite{Malshukov1997,Malshukov1999}, where $G(\theta )$ is only anisotropic, 
if both $\alpha ,\beta\neq 0$. Opposed to the narrow quantum wires considered where 
$\theta _\text{min}$, Eq.~(\ref{fullthetamin}), is a function of $\phi ,\alpha$ and $\beta$, 
in a 2DEG minimum of the conductivity appears either at $\theta =\pi /4$ or 
$3\pi /4$, depending on the sign of the product $\alpha \beta$, but independent of 
the ratio $\alpha /\beta$.\\
Apart from the condition $W\ll L^{\alpha /\beta}_\text{SO}$, the method should be applied at 
sufficiently small $B_{||}$ ($\lambda \ll 1$). As can be seen from Fig.~\ref{Fig3}b,c, when $\lambda \gtrsim 1$, 
$G$ is increased for any $\theta$, potentially changing the position of $\theta_\text{min}$ (see, e.g., blue triangles in Fig.~\ref{Fig3}c). Only for the case of $\alpha = \beta$ shown in Fig.~\ref{Fig3}a, $G(\theta _\text{min})$ does not increase, since $\vec{B}_\text{so}(\vec{k})\parallel\vec{B}_{||}$ for any $k$-vector. In this special case the validity of Eq.~(\ref{fullthetamin}) is not limited to narrow wires and small magnetic fields.

To summarize, in narrow quantum wires which exhibit weak localization even in the presence
of spin-orbit coupling, an in-plane magnetic field can suppress the weak localization effect.
We employed the unique angular dependence of this effect to suggest a method for the direct
and experimental determination of the ratio between Rashba- and Dresselhaus spin-orbit strengths 
from transport measurements. Its straightforward applicability may help to facilitate the 
design of semiconductor-based building blocks for spintronics.

{\em Acknowledgements}
We acknowledge valuable discussions with M. Wimmer, \.{I}. Adagideli and D. Bercioux. 
JN and MK acknowledge financial support from MEXT, 
MS from JSPS and the {\em Studienstiftung des Deutschen Volkes},
and KR from DFG through SFB 689.

%
%

\end{document}